\documentstyle[seceqn]{elsart}
\input epsf 
\begin{document}
\newcommand{\lesssim}{\,\vcenter{\hbox{$\buildrel\textstyle<\over\sim$}}}
\newcommand{\gtrsim}{\,\vcenter{\hbox{$\buildrel\textstyle>\over\sim$}}}

\begin{frontmatter}
\title{Critical Adsorption in Systems with Weak Surface Field:\\[4mm]
The Renormalization-Group Approach}
\author{Alina Ciach}
\address{Institute of Physical Chemistry, Polish Academy of Sciences,
Department III, Kasprzaka 44/52, 01-224 Warsaw, Poland}
\author{Uwe Ritschel}
\address{Fachbereich Physik, Universit\"at GH Essen, 45117 Essen, Germany\\[1cm]}
\maketitle
\begin{abstract} We study the surface critical behavior
of semi-infinite systems belonging to the bulk universality
class of the Ising model. 
Special attention is paid to the 
local behavior of experimentally relevant quantities such as
the order parameter and the correlation function
in the {\it crossover regimes} between different
surface universality classes, where
the surface field $h_1$ and the temperature enhancement $c$
can induce additional {\it macroscopic} length scales.
Starting from the field-theoretical
$\phi^4$ model and employing renormalization-group improved
perturbation theory ($\epsilon$ expansion),
explicit results for the local behavior of
the correlation and structure function (0-loop) and
the order parameter (1-loop) are derived. Supplementing earlier
studies that focussed on the special transition,
we here pay particular attention to the situation where a large $c$ suppresses
the tendency to order in the surface (ordinary transition)
but a surface field $h_1$ generates a
{\it small} surface magnetization $m_1$.
Our results are in good agreement with recent phenomenological
considerations and Monte Carlo studies devoted to similar
questions and, combined with the latter, provide a much
more detailed understanding
of the local properties of systems with weak surface fields.
\end{abstract}
\begin{keyword} Field theory, renormalization group,
surface critical phenomena, critical adsorption 
\end{keyword}
\end{frontmatter}

\newpage

\section{Introduction}\label{intro}
Surface critical phenomena are subject of increased current
experimental interest.
After the confirmation of some of the theoretical 
predictions  \cite{binder,diehl,didi,diwa}
in X-ray scattering experiments with FeAl \cite{mail},
the more recent efforts
focussed on binary mixtures near their consolute
point \cite{law,beysens,franck} and near-critical
fluids \cite{findenegg}. While for instance the
diffuse scattering of (evanescent) X-rays is governed
by the two-point correlation function,
in the experiments on fluids the
order parameter (OP), the concentration difference in fluid
mixtures or the density difference between liquid and gaseous
phase in single-component fluids,
plays a central role. For instance the reflectivity and
ellipticity
measured in light-scattering experiments are directly
related to the OP profile\cite{reflec,ellipt}.
Hence, a precise quantitative information about the local behavior
of these quantities is required to interpret
the experimental data.

In the framework of continuum field theory such as 
the $\phi^4$ model (belonging to the universality class
as the Ising model)
the surface influence is
taken into account by additional fields like the surface magnetic field
$h_1$ and the local temperature perturbation $c$ at $z=0$.
The latter can be
related to the surface enhancement of the spin-spin coupling in
lattice models \cite{diehl}.
At the bulk critical temperature,
$\tau\equiv(T-T_c^b)/T_c^b=0$, the tendency
to order near the surface
can be reduced ($c>0$), increased ($c<0$),
or, as a third possibility,
the surface can be critical as well (corresponding to
a particular value of $c$). As a result, each
bulk universality class in general divides into several distinct
{\it surface universality classes}, corresponding to the
fixed points of the renormalization-group flow and
labelled as ordinary ($ c \to \infty $),
extraordinary ($c \to -\infty $),
and special transition ($c=c^*$).

While a very well-developed theory for the individual 
universality classes exists, the picture in the crossover
regions between the fixed points is less complete. In particular
in view of some of the experiments, it may not be justified
to consider the system to be at a fixed point,
and so a detailed understanding of the
crossover regions is of special importance.  
Additionally the physics away from the fixed points is
much richer and more interesting as certain length scales emerge,
may become macroscopic, and compete
with the bulk correlation length, whereas with the surface fields
at their fixed-point values these scales are typically zero or infinity.

Near the special transition these phenomena
were studied already some time ago\cite{brezin,ciach}. It was realized
that $h_1$
gives rise to a length scale $l_{sp}\sim h_1^{-\nu/\Delta_1^{sp}}$ and
the singular behavior of thermodynamic quantities
goes through different universal regimes. This is
best illustrated by the spatial dependence of the
OP $m(z)$. For distances large compared to all length scales
it was known that $m(z)$ should decay
as $\sim z^{-\beta/\nu}$ \cite{binder}, where the exponent $\beta/\nu$
is the {\it scaling dimension} of the field $\phi$.  
This power law describes the asymptotic long-distance decay at $T_c^b$ when
the symmetry in the surface is broken spontaneously or explicitly,
i.e. at the extraordinary or normal transition, respectively\cite{bray,budi}.
For instance for the $d=3$ Ising
model $\beta/\nu\simeq 0.52$ \cite{ferlan}.  
For $z\ll l_{sp}$ (but still much larger than microscopic scales),
on the other hand, the result found in
Refs.\,\cite{brezin,ciach} was that $m(z)$ should behave as
$\sim z^{(\beta_1^{sp}-\beta)/\nu}$. Again, in the
three-dimensional Ising model the Monte Carlo literature
value for this combination of exponents is
$(\beta_1^{sp}-\beta)/\nu \simeq -0.15$ \cite{ruge}, whereas in
mean-field (MF) theory it is zero.
Thus, at the special transition, if fluctuation are taken into account,
the decay of $m(z)$ is still monotonic, however, it is described by
different power laws for $z\to 0$ and $z\to \infty$.

A similar phenomenon was recently pointed out
for the case of large $c$, i.e. a situation
near the ordinary transition\cite{czeri}.
For $c\gg c^*$ the
scaling field is given by ${\sf h}_1\equiv h_1/c^y$ \cite{diehl}
(where $y$ is a positive exponent to be discussed below in more detail)
and the length scale induced by this field is
$l_{ord}\sim {\sf h}_1^{-\nu/\Delta_1^{ord}}$.
The short-distance behavior of the magnetization is given by
$m\sim z^{\kappa}$ with $\kappa=(\Delta_1^{ord}-\beta)/\nu$.
This time the value of the short-distance exponent, $\kappa$,
is in general positive, such that the OP turns out
to be a {\it non-monotonic} function of $z$.
In the
Ising model $\kappa\simeq  0.21$, whereas
in MF theory, like at the special transition, one has $\kappa=0$.
At $z\approx l_{ord}$ again the crossover to
the ``normal" behavior $z^{-\beta/\nu}$ takes place. The near-surface growth
of order at the ordinary transition was recently also corroborated by Monte Carlo simulations\cite{swewa}.

Moreover, it was pointed out that
in a system with $c\gg c^*$ and non-vanishing $h_1$, it might be
the generic case that 
length scale $l_{ord}$ becomes macroscopic and comparable
or even larger than the bulk correlation length $\xi$\cite{alina}. The
behavior of thermodynamic quantities near the surface is then
governed by exponents of the ordinary transition. 
As a consequence, for $\xi \simeq l_{ord}$
the amount of adsorbed order as a function of $\tau$ is described
by the power law $\tau ^{\beta-\Delta_1^{ord}}$ \cite{alina},
in very good agreement with experiments \,\cite{findenegg,franck}.
Only much closer to the critical point a crossover to $\tau^{\beta-\nu}$
takes place, as expected for the normal transition and
assumed in previous studies.

The main purpose of our present work is a careful derivation of
the OP scaling function for the crossover between ordinary and
normal transition in the framework of renormalization-group
improved perturbation theory.
While the respective studies in Ref.\,\cite{czeri}
focussed on the near-surface behavior, and the Monte Carlo simulations of
Ref.\,\cite{swewa} were hampered by strong finite-size effects
and did not yield OP profiles suitable for the comparison with
experimental data, the results presented below give to
one-loop order the OP profiles for the semi-infinite system.
Further we present MF results for the structure function, relevant
for scattering of X-rays for example, in the regime between
ordinary and normal transition. Eventually we discuss the crossover
between special and ordinary transition that (while probably
less important in view of experiments) is an interesting
and technically demanding problem in perturbation theory.     

The rest of this paper is organized as follows: In Sec.\,\ref{background}
we discuss the behavior of thermodynamic quantities from the viewpoint
of a scaling analysis, give an heuristic argument for the
growth of order at the ordinary transition, and compare with
the situation in critical dynamics where analogous phenomena occur
after temperature quenches.
In Sec.\,\ref{0loop}, as the building blocks for the perturbative
calculation, we present the results of the MF (zero-loop) theory. Especially
the correlation function for general $c$ and $h_1$
is discussed in some detail, and
the structure function for critical diffuse scattering
is calculated numerically from that result.
In Sec.\,\ref{1loop} the OP profiles are calculated
to one-loop order, and special emphasis is put on
the crossovers between ordinary and normal transition and special
and ordinary transition, respectively.

\section{Background}\label{background}

\subsection{Scaling Analysis for the Order Parameter}\label{scalan}
In this section we study the behavior of thermodynamic variables
with the help of a phenomenological scaling analysis. As the
most instructive example, we discuss the local behavior
OP for different surface universality classes
and the various crossovers. Certain aspects of the scaling behavior
of the correlation function will be treated in Sec.\,\ref{0loop}.

Since the parameter $c$ has to be fine tuned in addition to
$\tau$, from the viewpoint of the experimentalist the special
transition is certainly the 
most exotic among the surface universality classes.
However, from the viewpoint of the
scaling analysis it is
the most straighforward case. 
Near the special transition $h_1$ and $\tilde c\equiv c-c^*$ are the linear
scaling fields pertaining to the surface.  
In the critical regime thermodynamic quantities are described
by homogeneous functions of the scaling fields. 
Let us consider the OP for
small $h_1$ and $\tilde c>0$. Its
behavior under
rescaling of distances should be described by
\begin{equation}\label{scalsp1}
m(z,\tau,{h}_1,c)\sim b^{-\beta/\nu}\,m(z\,b^{-1},\,\tau b^{1/\nu},\,
{h}_1\,b^{\Delta_1^{sp}/\nu},\,\tilde c \,b^{\Phi/\nu})\>,
\end{equation}
where all exponents have their standard meaning\cite{binder}.
In general the surface exponent $\Delta_1$ occurring in
(\ref{scalsp1}) has different values
for different surface universality classes\cite{binder}, and
therefore it has been marked by `{\it sp}' (for special).

Removing the arbitrary rescaling parameter $b$ in
Eq.\,(\ref{scalsp1}) by setting it
$\sim z$, one obtains the scaling form of the magnetization
\begin{equation}\label{scalspm}
m(z,\tau,h_1)\sim z^{-\beta/\nu}\,{\cal M}(z/\xi, z/l_{sp}, z/l_c)\>,
\end{equation}
where
\begin{equation}\label{lengthsp}
l_{sp}\sim h_1^{-\nu/\Delta_1^{sp}}
\qquad\mbox{and}\qquad l_c\sim \tilde c^{-\nu/\Phi}
\end{equation}
are the length scales determined by $h_1$ and $c$, respectively.
The other length scale pertinent to the semi-infinite system and occurring in
(\ref{scalspm}) is the bulk correlation length $\xi\sim\tau^{-\nu}$.

In order to discuss the various asymptotic cases, let us
set $\tau=0$ for simplicity. In other words, we assume for the
moment that the bulk correlation length is much larger than
any length scale set by the surface fields, in which limit
the scaling function ${\cal M}$ in (\ref{scalspm})  becomes
\begin{equation}\label{scalc}
{\cal M}(z/\xi,z/l_{sp}, z/l_c)\approx {\cal M}_{crit}(z/l_{sp},z/l_c)\>.
\end{equation}
Further below we will discuss modifications due to finite $\xi$. 
Let us first consider the case $\tilde c=0$ where both
bulk and surface are in the critical state at $T_c^b$. 
Then the only remaining length scale is
$l_{sp}$ and
the OP profile can be written in the
scaling form
\begin{equation}\label{h1}
m(z,{h}_1)\sim z^{-\beta/\nu}\,{\cal M}_{sp}(z/l_{sp})\>.
\end{equation}

As mentioned in the Introduction,
for $z\to \infty$ the magnetization decays as $\sim
z^{-\beta/\nu}$
and, thus, ${\cal M}_{sp}(\zeta)$ should approach a constant for
$\zeta\to \infty$.
In order to work out the {\it short-distance}
behavior,
we demand that $m(z)\sim m_1$ when $z$ is small (but still
larger than microscopic scales). This
is motivated by and consistent with the
field-theoretic short-distance expansion \cite{syma,diehl}, where,
on a more fundamental level,
{\it field operators} near a boundary are represented in terms of
boundary operators multiplied by $c$-number functions.

Now, the dependence of $m_1$ on $h_1$ is given by
$m_1 \sim h_1^{1/\delta_{11}^{sp}}$, and with the
scaling relation $\delta_{11}=\Delta_1/\beta_1$ we obtain that
the scaling function ${\cal M}_{sp}(\zeta)$ in (\ref{h1}) should
behave as $\sim \zeta^{\beta_1^{sp}/\nu}$ in the limit
$\zeta\to 0$. Inserting this in (\ref{h1}) leads
to the short-distance behavior
\begin{equation}\label{shodisp}
m(z)\sim h_1^{1/\delta_{11}^{sp}}\,z^{(\beta_1^{sp}-\beta)/\nu}
\qquad \mbox{for}\qquad z\ll l_{sp}\>,
\end{equation}
in agreement with Refs.\,\cite{brezin,ciach}.
In other words: Near the special transition a small $h_1$ gives rise
to a macroscopic regime near the surface of depth $l_{sp}$
on which the behavior of the OP is governed
by an exponent different from the one describing the
long distance behavior for $z\gg l_{sp}$. For the
three-dimensional Ising model the short-distance exponent
$(\beta_1^{sp}-\beta)/\nu$ is negative, its
(MC) literature value being $-0.15$ \cite{ruge}
(compared with $-\beta/\nu \simeq -0.52$ \cite{ferlan}
that governs the decay for $z\gg l_{sp}$) .

Next let us consider the case of {\it large} $c$, near the
ordinary transition, the situation which is more natural
if one thinks of possible applications for experiments and
which was discussed in Ref.\,\cite{czeri}.
Since $c$ is a so-called
{\it dangerous irrelevant variable} it must not be simply set
to $\infty$ from the start\cite{diehl}. A careful analysis reveals that
close to the ordinary transition the linear scaling field
is given by ${\sf h}_1 = h_1/c^y$ with $y=(\Delta_1^{sp}-
\Delta_1^{ord})/\Phi$. In MF theory the value of
the exponent $y=1$. Further, it is discussed
in detail in Ref.\,\cite{diehl} that in the framework of the $\epsilon$
expansion one does not capture the deviation from the MF value
in this exponent, while e.g. the $z$-dependence 
of expectation values is reproduced correctly. 
  
Analogous to (\ref{scalsp1}), also near the ordinary transition the magnetization is a homogenous function of the linear scaling fields:
\begin{equation}\label{scalord1}
m(z,\tau,{\sf h}_1)\sim b^{-\beta/\nu}\,m(z\,b^{-1},\,\tau b^{1/\nu},\,
{\sf h}_1\,b^{\Delta_1^{ord}/\nu})\>.
\end{equation}
Removing $b$ by setting it
$\sim z$ and setting $\tau=0$ again, one obtains 
the scaling form
\begin{equation}\label{scalordm}
m(z,{\sf h}_1)\sim z^{-\beta/\nu}\,{\cal M}_{ord}(z/l_{ord})\>,
\end{equation}
where
\begin{equation}\label{lengthord}
l_{ord}\sim {\sf h}_1^{-\nu/\Delta_1^{ord}}
\end{equation}
is in this case the only length scale determined by the surface fields.

In order to analyze the short-distance behavior of the
magnetization, we demand again that $m(z\to 0) \sim m_1$. 
Since for large $c$ the surface is paramagnetic and responds
linearly to a small external field, we have now $m_1\sim {\sf h}_1$ \cite{bray}.
The immediate consequence of the simple {\it linear} response
for the scaling function ${\cal M}_{ord}(\zeta)$ occurring in (\ref{scalordm})
is that it has to behave as $\sim \zeta^{\Delta^{ord}_1/\nu}$ for
$\zeta\to 0$. Inserting this in (\ref{scalord1}),
we obtain that for $z\ll l_{ord}$ the
magnetization is described by
\begin{equation}\label{power}
m(z)\sim {\sf h}_1 \,z^{\kappa}\qquad \mbox{with}\qquad \kappa\equiv
\frac{\Delta_1^{ord}-\beta}{\nu}
\end{equation}
Using that $\Delta_1/\nu=(d-\eta_{\parallel})/2$ and $\beta/\nu=
(d-2+\eta)/2$ together with the
scaling relation $\eta_{\perp}=(\eta+\eta_{\parallel})/2$, 
one obtains $\kappa=1-\eta_{\perp}^{ord}$.

In the MF approximation
the result for $\kappa$ is zero. 
However, a positive value and, consequently, a 
non-monotonic profile are obtained when fluctuations
are taken into account
below the upper critical dimension $d^*=4$.
The numeric value of $\kappa$ is 0.21, where
we have used the Monte Carlo literature value $0.8$
for $\beta_1^{ord}/\nu$ \cite{ruge2} together with the scaling relation
$\Delta_1+\beta_1=(d-1)\nu$\cite{binder}.

\subsection{Crossover between Special and Ordinary Transition}\label{crospord}
In the previous section we discussed the behavior
of the OP in situations where $c=c^*$ and $c\to \infty$
and where $h_1\neq 0$.
In both cases we found a near-surface behavior governed by
the exponents of the special or ordinary transition, respectively,
and a crossover to the power law $z^{-\beta/\nu}$ characteristic
for the normal transition.

However, the scaling form (\ref{scalspm}) should also
describe the behavior of $m(z)$ in the general
case, in the presence of two length scales, and, in particular,
it should cover the
crossover between ``special'' and ``ordinary'' near-surface behavior.
Let us assume that $h_1$ is fixed and vary $c$ between $c^*$ and $\infty$
such that the length scale $l_c$ ranges from $\infty$ and $0$.
First, in the limit $l_c\to \infty$ we were lead from (\ref{scalc}) to
(\ref{h1}).
Due to the fact that the linear scaling field at the ordinary transition is a combination
of the surface parameters $h_1$ and $c$,  the second limit, $z/l_c\to \infty$, is more subtle. In this limit the scaling
function ${\cal M}_{crit}$ in (\ref{scalc}) should behave as
\begin{equation}\label{lim2}
{\cal M}_{crit}(z/l_{sp},z/l_c \to \infty)\approx {\cal M}_{ord}((z/l_{sp})^a\cdot (z/l_c)^{1-a})\>,
\end{equation}
where $a=\Delta_1^{sp}/\Delta_1^{ord}$ such that the product in the
argument of ${\cal M}_{ord}$ on the right-hand side of (\ref{lim2}) is just $z/l_{ord}$.

Although experimentally a system where
both scales $l_{ord}$ and $l_c$ are macroscopic
is probably not realizable, in the
field-theoretic model to be analyzed below there
is no principle limitation.
So at least from the theoretical point
of view it is an interesting question to ask
what shape the OP profiles have in the crossover
regime between special and ordinary transition. 
In the following we present a qualitative scenario that,
to our present knowledge, describes
this crossover.

Consider a system with fixed $h_1$ and variable $l_c$.
As long as $l_c>0$, the asymptotic
behavior for small distances, $z\ll l_c$, is described by (\ref{shodisp}) where
the amplitude of the short-distance power
law does not depend on $c$. On the other hand, for distances
large compared to all scales induced by $h_1$ and $c$, the OP
profile should behave as
$\sim z^{-\beta/\nu}$,
again with an amplitude independent of $c$
(and $h_1$). More precisely, the crossover to the ``normal"
behavior takes place at a
distance $l_{int}$ that varies between
$l_{sp}$ for $c=c^*$ and $l_{ord}$ for large $c$.

The qualitative shape of the crossover
profiles is depicted in Fig.\,1. As long as $l_c>l_{int}$, the
OP is a monotonic function of $z$, with
a single crossover at $z=l_{int}\simeq l_{sp}$.
For $l_c<l_{int}$, however,
the profile becomes non-monotonic and exhibits
several crossovers. For $z>l_c$ it first drops
and thereafter increases again to approach the
long-distance power law. For $l_c\to 0$ eventually the
increase for $z < l_{int}\simeq l_{ord}$ must approach
the power law (\ref{power}). Of course, in this limit
it would be appropriate to rescale the $z$-axis or readjust $h_1$ such
that $l_{ord}$ remains finite.  
\begin{figure}[t]
\def\epsfsize#1#2{0.6#1}
\hspace*{1.8cm}\epsfbox{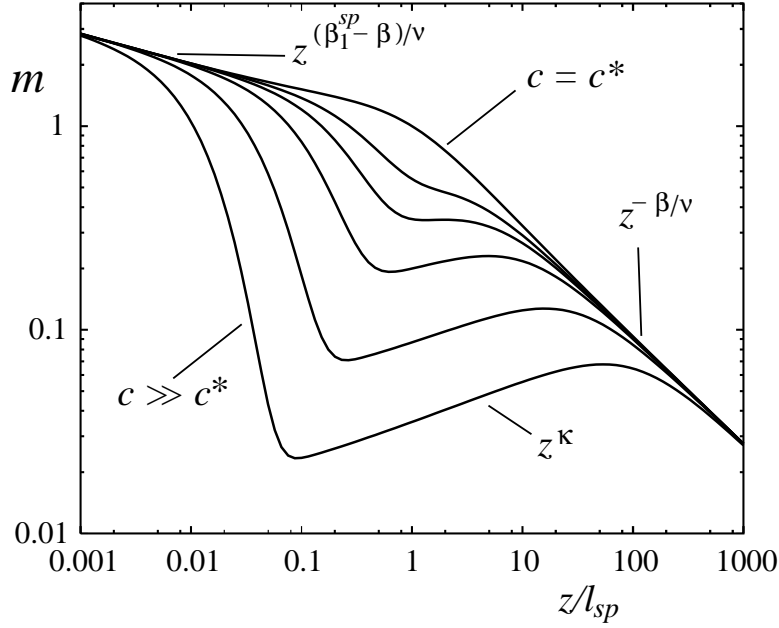}
\caption{Qualitative scenario for the crossover between special
and ordinary transition for the order parameter. In our representation
$h_1$ is fixed and $c$ varies from $c=c^*$ to a value $c\gg c^*$.}
\end{figure}

\subsection{The Correlation Function between Ordinary and Normal Transition}\label{correl}
As for the OP in the previous sections, a similar scaling
analysis combined with arguments taken from short-distance
expansions can also be carried out for the correlation function.
Here we restrict ourself to a qualitative discussion of
the two-point correlation function in directions parallel to the
surface, called parallel correlation function
in the following, for a system with large $c$.
This function is useful to illustrate the general features
one has to expect for correlation functions in the
crossover regimes and later on it is needed
for an heuristic argument that explains the near-surface growth
expressed in (\ref{power}). Further, we restrict the discussion
again to the case $\tau=0$.

\begin{figure}[h]
\def\epsfsize#1#2{0.6#1}
\hspace*{2cm}\epsfbox{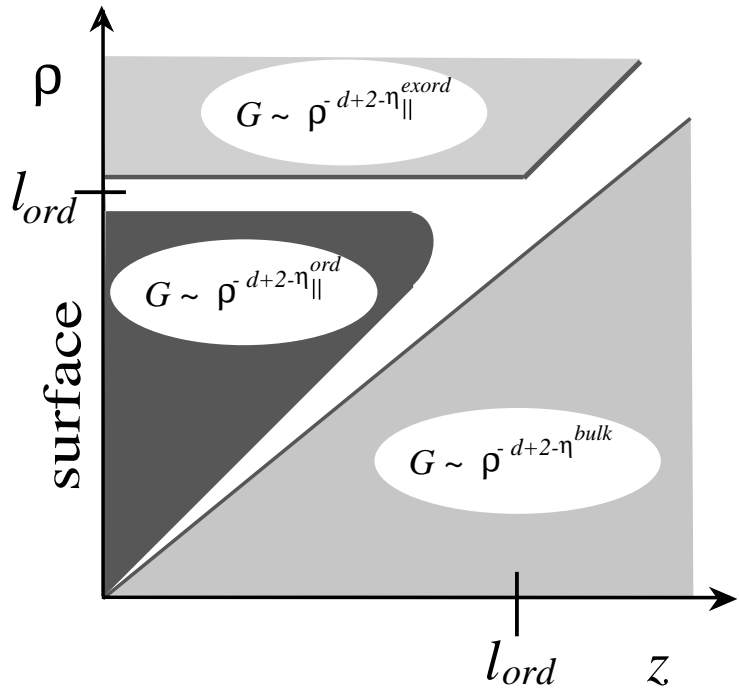}
\caption{Regimes of universal behavior (gray areas)
of the correlation function in directions parallel to the surface.
The respective power laws governing the $\rho$-dependence of
$G(\rho,z)$ are shown.
Crossover regimes are represented as white areas.}
\end{figure}

The different regimes which the
parallel correlation function
$G(\rho,z,z)$ (where $\rho$ denotes the distance between
two point in a plane parallel to the surface) goes through are depicted in
Fig.\,2. First of all, for $\rho < z$ the function $G$ asymptotically
behaves as in the bulk, $G$ is governed by the bulk exponent $\eta$
(lower light gray area).
In the opposite case, $\rho> z$, the result depends on whether
the distance from the surface is smaller or larger than the
length scale $l_{ord}$ introduced in (\ref{lengthord}).
In the case $z> l_{ord}$ for $\rho\simeq z$ the crossover
from bulk to surface behavior takes place, where $G$
is governed by the anomalous dimension of the
normal transition (upper light gray area). The situation for $z<l_{ord}$
is different. Again at $\rho\simeq z$ a crossover to surface behavior
takes place, to a regime $z< \rho <l_{ord}$ where the behavior
is governed by exponents of the {\it ordinary} transition (dark gray area).
Eventually, for $\rho \simeq l_{ord}$ the crossover to the regime
with ``normal" behavior takes place. In Fig.\,2 the crossover
regions are represented as white areas.

As we shall see in Sec.\,\ref{scatter}, this crossover behavior
leads to the result that for instance in scattering
with finite penetration depth of the radiation
and $l_{ord}$ greater than this depth,
the structure function is effectively
governed by exponents of the ordinary transition.      

\subsection{Modifications at $T\neq T_c^b$}\label{temp}
The scaling analysis 
presented above can be straightforwardly extended
to the case $\tau>0$. In $d>2$, we may assume that the
behavior near the surface for $z<<\xi$ is unchanged compared
to (\ref{shodisp}) and (\ref{power}).
The behavior farther away from the surface depends
on the ratio $l/\xi$, where $l$ stands
for $l_{sp}$ or $l_{ord}$. 
In the case of $l>\xi$ a crossover
to an exponential decay will take place for $z\simeq \xi$
and the regime of nonlinear decay $\sim z^{-\beta/\nu}$
does not occur.
For $l < \xi$ a crossover
to the power-law decay $\sim z^{-\beta/\nu}$ takes place
for $z\simeq l$
and finally at $z\simeq \xi$ the exponential behavior sets in.

Similar results hold for the correlation functions. As long
as for example the distance $\rho$ in the parallel
correlation function is smaller than $\xi$, the discussion
of Sec.\,\ref{correl} holds. For distances larger than $\xi$
correlations decay exponentially.

An interesting phenomenon can be observed in the case
$\xi <l$. As said above,
$m(z)$ then never reaches the regime with power-law decay, but
crosses over from the near-surface regime directly to the exponential
decay. Let us restrict the further discussion to the 
case of the ordinary transition.
Since the region where $m(z)$ grows then only extends up to the
distance $\xi$, the maximum of the OP profile
has roughly the
value $m_{max} \simeq {\sf h}_1\,\xi^{\kappa}$, where
the exponent $\kappa$ was defined in (\ref{power}). Now,
the amplitude of the exponential decay should behave
as $\sim m_{max}$ such that for $z\gg \xi$ we have 
\begin{equation}\label{expo}
m(z)\sim{\sf  h_1}\,\xi^{\kappa}\,\exp(-z/\xi)\>.
\end{equation}
In other words, in the case $\xi < l_{ord}$ the exponent
$\kappa$ not only governs the behavior of $m(z)$ near the surface,
but also leaves its fingerprint much farther away
of an universal dependence of the amplitude of the
exponential decay.
Nothing comparable occurs when $\xi =\infty$
(compare Sec.\,\ref{scalan} above), where
all profiles approach the same
curve $m(z)\approx {\cal A}\, z^{-\beta/\nu}$
for $z\gtrsim l_{ord}$, with an amplitude ${\cal A}$ independent
of $h_1$. A similar phenomenon, termed ``long-time memory'' of
the initial condition, does also occur in critical
dynamics for $T\ge T_c^b$ \cite{own2}. The analogies between surface
critical phenomena and critical dynamics will be discussed in more
detail in the next section.

\subsection{Relation to Critical Dynamics and Heuristic Arguments}\label{critdyn}
The {\it spatial} variation of the
magnetization discussed so far, especially the increasing
profile at the ordinary transition,
strongly resembles the {\it time} dependence of the
OP in relaxational processes at the critical point.
If a system with nonconserved OP (model A) is quenched
for example from a
{\it high-temperature initial state} to the critical point, with a small
initial magnetization $m^{(i)}$, the OP behaves as $m \sim
m^{(i)}\,t^{\theta}$ \cite{jans}, where the short-time exponent $\theta$
is governed
by the difference between the scaling dimensions of initial
and equilibrium magnetization divided by the
dynamic (equilibrium) exponent\cite{own1}. Like the exponent $\kappa$
in (\ref{power}), the exponent
$\theta$ vanishes in MF theory, but becomes positive
below $d^*$. For example, its value in the
three-dimensional Ising model with Glauber
dynamics is $\theta=0.108$ \cite{theta}.

The high-temperature initial state of the relaxational process is to
some extent analogous
to the surface that strongly disfavors the order and that (for
$h_1=0$) belongs to the universality class of the ordinary transition.  
Further expanding this analogy, heating a system
from a low-temperature (ordered) initial state to
the critical point would be similar to the situation
at the extraordinary transition. Eventually, analogous to
the special transition would be a ``relaxation process'' that starts
from an equilibrium state at $T_c^b$. 

Motivated by the analogy between surface
critical phenomena and critical dynamics,
an heuristic argument for the short-distance growth of the OP at the ordinary transition behavior can be brought forward.
In relaxation processes, after the system has passed through
a non-universal regime immediately after the quench, the growth
of correlation length $\xi(t)$ is described by a power law.
Analogously one may
argue that in the static case in the surface-near regime there
exists an effective parallel correlation length $\xi_{\parallel}(z)$
that grows as $\sim z$ when the distance $z$ grows away from the surface.
As discussed in Sec.\,\ref{correl},
for lateral distance $\rho\simeq z$ the parallel correlation
function crosses over from ``bulk'' to  ``surface decay", the latter being
much faster than the former (compare Sec.\,\ref{scatter}).

For the sake of the argument, 
consider the semi-infinite Ising system
at bulk criticality where initially $h_1=0$ with the tendency
to order in the surface reduced compared with the bulk. 
If then a small $h_1$ is imposed on the surface, a number of 
spins will change their orientation to generate a small $m_1$.
Now, the magnetization induced by
in a distance $z$ from the surface is
determined by several factors. 
First, it is proportional to $m_1$ and, in turn, 
for small $m_1$, it is proportional to $h_1$.
Second, it should be proportional to the {\it perpendicular} correlation
function $C(0,z)\sim z^{-d+2-\eta^{ord}_{\perp}}$ which
says how much of a given surface configuration ``survives" in a distance $z$.
Eventually, it should be proportional to the
correlated area in the distance $z$ that is influenced
by a single surface spin.
According to the discussion above, the latter
grows as $\xi_{\parallel}^{d-1}\sim z^{d-1}$.
Taking these factors together we obtain
\begin{equation}\label{heuristic}
m(z)\sim h_1 \,C(0,z)\, \xi_{\parallel}^{d-1}\sim h_1\,z^{1-\eta^{ord}_{\perp}}\>,
\end{equation}
in agreement with the short-distance law (\ref{power}).

The above argument holds as long as $z$ and $m_1$ are small
enough. If $z$ is larger than the average
distance between the spins flipped,
the correlated areas start to overlap and the above argument
breaks down. This is where the crossover to the normal
transition sets in.

From the field-theoretical point of view in both cases
(surface critical phenomena and critical dynamics) 
the modified power laws are due to additional short-distance
or short-time singularities near spatial or temporal
boundaries. In both cases the singular behavior near the surface
can be described by means of short-distance expansions or
(in simplified form) by means of a phenomenological
scaling analysis as presented above.
It is interesting to point out, however, that in the presence of both spatial
{\it and} temporal boundaries no additional singularities
(other than the ones discussed above) occur \cite{own3}, and,
as a consequence,
the {\it short-time behavior in critical dynamics near surfaces}
is governed by exponents that can be expressed in terms of
known surface and dynamic exponents\cite{own3}.

\section{Mean-field (Zero-Loop) Approximation}\label{0loop}
\setcounter{figure}{2}
In this section we define the model and
describe the results of the MF theory for
general $c$ and $h_1$. They are the building blocks for
the (one-loop) perturbative calculation of Sec.\,\ref{1loop}.
Especially the zero-loop propagator for general $c$ and $h_1$
is not contained in explicit form in the literature so far.
Further in Sec.\,\ref{scatter} we discuss the behavior of the
real-space correlation function and the respective structure function
that is measured for instance in X-ray scattering experiments.
The one-loop calculation for this quantity for general $c$ and
$h_1$ is not attempted in the present work. So the MF results presented
here are not only the best presently available in the framework
of continuum field theory, but also provide, as we think, a
qualitative description of the results
to be expected from scattering experiments. 

\subsection{Model}\label{Model and Mean-Field Order Parameter}  
We consider the semi-infinite scalar $\phi^4$ with the Hamiltonian\cite{diehl}
\begin{equation}\label{hamiltonian}
{\cal H}=\int_{V} \left[\frac12 (\nabla \phi)^2+\frac12 \tau\, \phi^2
+\frac{1}{4!}u\,\phi^4\right] +
\int_{\partial V} \left[\frac12 c \,\phi^2+h_1\,\phi
\right]\>,
\end{equation}
where $V$ and $\partial V$ stand for the volume of the
semi-infinite system and its (planar) surface, respectively.

The well-known solution to the OP profile,
minimizing the above Hamiltonian, takes the form\cite{lubrub,bray,brezin}
   \begin{equation}\label{mMF}
   m^{(0)}(z) =\alpha^{-1}(z+z^+)^{-1}\>,
   \end{equation}
where the length scale $z^+$ is given by
  \begin{equation}\label{z0}
  z^+=\frac{c+\sqrt{c^2+4h_1 \alpha}}{2\alpha h_1}
  \end{equation}
and  
  \begin{equation}\label{alpha}
  \alpha=\sqrt{\frac{u}{12}}\>.
  \end{equation}
The solution (\ref{mMF})
satisfies the general (mixed) boundary condition
\begin{equation}\label{boundco}
\frac{\partial\,m}{\partial z}=c\,m-h_1\>.
\end{equation}

The MF solution to the OP profile has a fairly simple form.
Consistent with the discussion in Secs.\,\ref{intro}
and \ref{background}, it is a monotonic function of $z$.
Further, although $c$ and $h_1$ in
principle give rise to two independent
length scales, in the OP profile they show up
in a single length scale $z^+$ only.
As a consequence, the shape of the profile
does not change qualitatively when one moves from one fixed point
to the other.

\subsection{Free Propagator}\label{geom}
The two-point correlation function
for general $c$ and $h_1$ that will be described next
solves the equation
   \begin{equation}\label{Geq}
   \left[-\nabla^2 +\frac{1}{2} \,u\, m^2(z)\right]G({\bf x},
   {\bf  x}')=\delta({\bf x}-{\bf x}')\>,
   \end{equation}
subject to the boundary condition (at $z=0$)
   \begin{equation}\label{bound}
   \left[\frac{\partial}{\partial z}-c\right]G({\bf x},{\bf x}')|_{z=0}=0\>,
   \end{equation}
where $m^{(0)}(z)$ is the profile of (\ref{mMF}).
To calculate $G$ we Fourier transform
with respect to the parallel, translationally invariant spatial
directions. The result that can be obtained by standard methods
reads
  \begin{equation}\label{prop}
  \hat G({\bf p};z,z') =\theta(z-z')W(p,z)U(p,z')+\theta(z'-z)W(p,z')U(p,z)\>,
  \end{equation}
where $p=|{\bf p}|$, $\theta(z)$ is the Heaviside step function, and
the function $W$ and $U$ are given by
  \begin{equation}\label{W1}
  W(p,z) =\left[p^2+\frac{3p}{z+z^+} +\frac{3}{(z+z^+)^2}\right]e^{-pz}
  \end{equation}
and
   \begin{eqnarray}\label{U}
   U(p,z)&=&\frac{1}{2p^5}\left\{ B(p)\left[p^2+
   \frac{3p}{z+z^+}   +\frac{3}{(z+z^+)^2}
 \right]e^{-pz} \right.\nonumber\\[3mm]
  & &\hspace*{2.5cm}  \left.+  \left[p^2-\frac{3p}{z+z^+} +\frac{3}{(z+z^+)^2}\right]e^{pz}\right\}
   \end{eqnarray}
with
    \begin{equation}\label{BB}
    B(p)=\frac{k^3-k^2(q+1)+3kq-3q}{k^3+k^2(q+1)+3kq+3q}\>.
    \end{equation}
Further, we introduced a dimensionless wave vector
   \begin{equation}\label{k}
   k=p\,z^+
   \end{equation}
and, for the sake of brevity, we defined
  \begin{equation}\label{q}
  q=2+c\,z^+\>.
  \end{equation}
The length scale $z^+$ was already introduced in (\ref{z0}).

In contrast to
the OP profile,
the MF correlation function now depends on two length
scales. In addition to the dependence on $z^+$, it contains via
the quantity $q$ defined in (\ref{q}) an explicit
$c$-dependence. The latter defines the second length scale
$l_c\equiv c^{-1}$. 

For the correlation function $\hat G({\bf p};z,z) $ and
the for the propagator at ${\bf p}=0$, one obtains from (\ref{prop})
   \begin{eqnarray}\label{chi}
   \hat G({\bf p};z,z) &=&\frac{1}{2p^5}\left\{ B(p)\left[p^2+
   \frac{3p}{z+z^+}  +\frac{3}{(z+z^+)^2}\right]^2 e^{-2pz}\right.\nonumber\\[3mm]
& & \left. \hspace{2.5cm}+ p^4  -\frac{3p^2}{(z+z^+)^2} +\frac{9}{(z+z^+)^4}\right\}
   \end{eqnarray}
and
  \begin{eqnarray}\label{G(0)}
  \hat G(0;z,z') &=&\left.\frac{1}{5}\;\right\{ 
  (z+z^+)^{-2}\times \nonumber\\
& &\hspace*{-1.4cm}\left.   \left[\frac{5-q}{q}\,(z^+)^5\,(z'+z^+)^{-2}+
  (z'+z^+)^3\right] \theta(z-z') + (z\leftrightarrow z')\right\}\>,
  \end{eqnarray}
respectively.
The results (\ref{chi}) and (\ref{G(0)})
are needed in Sec.\,\ref{1loop}
to calculate the one-loop contributions to the OP profile.

\subsection{Structure Function for Critical Diffuse Scattering}\label{scatter}
A quantity directly relevant for experiments is the structure function
\begin{equation}\label{scatint}
\tilde S(p,\kappa) = A\,\int_0^{\infty}\int_0^{\infty}
\hat G({\bf p},\,z,\,z')\, e^{i(\kappa z-\kappa^*z')}\,dz\,dz'\>,
\end{equation}
where $A$ is a constant -- for a cubic lattice it is just the
illuminated area divided by a power of the lattice
constant -- and $\kappa=\kappa_r+i\,\kappa_i$ is a complex wavenumber
taking into account that the radiation penetrates the material
roughly up to the depth $\kappa_i^{-1}$ \cite{diwa}.

Analytic results for $\tilde S$ on the basis of
(\ref{prop}) can be obtained at or near the fixed points only, where
the correlation function turns out to be much
simpler than the general expression
(\ref{prop}). Here we present
the analytic results
for the ordinary and the extraordinary (or normal)
transition and evaluate the transformation
(\ref{scatint}) numerically in the
crossover regime, i.e. for large $c$ and
and arbitrary $h_1/c$. The latter is the generic situation
met in systems where the spin-spin interaction near
the surface is not enhanced and where the surface spins are coupled
to an external (non-critical) medium that favors one of the
spin directions. The results presented below were obtained by
setting $q=\infty$ in (\ref{prop}) (thus ignoring
corrections of order $c^{-1}$). Then
the function $B(p)$ becomes
\begin{equation}
\lim_{c\to \infty}\,B(p)=-\frac{k^2-3k+3}{k^2+3k+3}\>,
\end{equation}
and the length scale $z^+$ from (\ref{z0})
is to be identified with the length scale $l_{ord}$
introduced in Sec.\,\ref{scalan}. Again in leading order
in $c^{-1}$ one obtains
\begin{equation}\label{lord}
l_{ord} =z^+ = \frac{c}{\alpha h_1}\>,
\end{equation}
consistent with the scaling analysis in Sec.\,\ref{scalan}
and with the fact that $y=1$ in MF theory.

By general arguments (scaling analysis and short-distance expansion)
it was shown by Dietrich and Wagner\cite{diwa} that
in the limit $p\to 0$ the structure function takes the form
\begin{equation}\label{critdif}
\tilde S(p,\kappa) = s_0(\kappa)-s_1(\kappa)\,p^{\eta_{\parallel}-1}\>.
\end{equation}
Explicit results were obtained in Ref.\,\cite{diwa} at the 
ordinary transition up to one-loop order.

Starting from (\ref{prop}) our result for the ordinary transition
($\tau=h_1=0,\,c=\infty$) is
\begin{equation}\label{strucord}
A^{-1}\,\tilde S(p,\kappa)=\frac{1}{2\,\kappa_i\,|\kappa|^2}-
\frac{p}{|\kappa|^4}\>,
\end{equation}
in consistency with Ref.\,\cite{diwa}.
For the normal transition ($\tau=0,\,c=h_1=\infty$) we obtain 
\begin{equation}\label{strucnorm}
A^{-1}\,\tilde S(p,\kappa)=\frac{1}{20\,\kappa_i^3}\,\Re\left[ _2\hspace{-0.1mm} F_1
(1,3,5,i\,\kappa_i^{-1}\kappa/2)\right]-\frac{12}{75}\,
\frac{p^5}{|\kappa|^8}\>,\end{equation}
where $_2\hspace{-0.1mm} F_1$ is the hypergeometric function \cite{grad}
and $\Re$ denotes the real part of the term in square brackets.
The MF values of $\eta_{\parallel}$ are $2$ for the ordinary transition
and $6$ for the normal transition, respectively,
and so the results (\ref{strucord}) and (\ref{strucnorm}) are
consistent with (\ref{critdif}). 

\begin{figure}[b]
\def\epsfsize#1#2{0.6#1}
\hspace*{2cm}\epsfbox{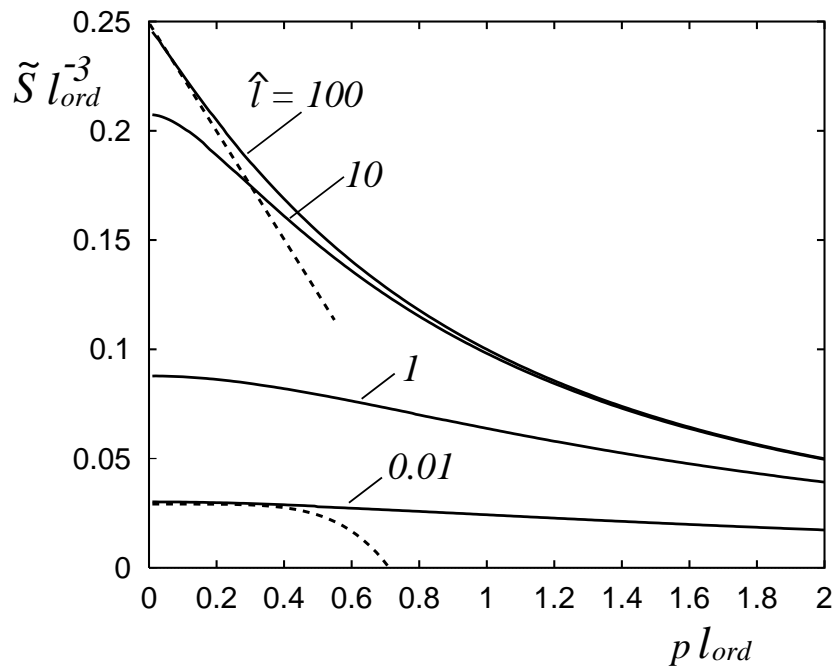}
\caption{The MF structure function for critical diffuse scattering defined in
(3.16) in the crosover regime between ordinary and normal transition
for different values of $l_{ord}\,\kappa_i$ (solid lines). The upper dashed
line shows the asymptotic ($p\to 0$) result at the ordinary transition given
in (3.20). The lower dashed line is the analogous curve at the
normal transition of (3.21).}
\end{figure}

As the next step the transformation (\ref{scatint}) is carried out
numerically in the crossover regime between ordinary and normal
transition. The result can certainly not be compared with
experiments on a quantitative
level -- the real exponents differ significantly from their MF
values -- , but it should give the qualitative picture that one has to
expect from a (Gedanken)experiment carried out in a system 
at bulk criticality with adjustable scaling field $h_1/c$
and otherwise fixed parameters.

The results for $\tilde S$ for the particular choice
$\kappa_r=\kappa_i$ are depicted in Fig.\,3 (solid lines).
For $\hat l \equiv l_{ord}\cdot\kappa_i\gg 1$, the $p\to 0$
limit of $\tilde S$ has effectively the cusp
shape typical for ordinary transition. 
(In contrast to the situation in the bulk, $\tilde S$ never diverges\cite{diwa},
because in the surface-near regime (which at $T_c^b$ extends to
infinity) the long-distance decay of correlations
is much faster than in the bulk.) For smaller values of $\hat l$,
it becomes visible that for small momenta $p\,l_{ord}\ll 1$
the behavior is actually described by the anomalous dimension
of the normal transition, leading to a curve without cusp 
departing with slope zero from $p=0$. 
For comparison we have also plotted the asymptotic results for
$p\to 0$ and $l_{ord}=0$ and $\infty$ (dashed lines).

\section{One-Loop Calculation of the Order-Parameter Profile}\label{1loop}
\setcounter{figure}{3}
In this section we calculate the OP profile in the framework
of the $\epsilon$-expansion (where $\epsilon\equiv 4-d$)
near both the ordinary and special fixed point,
the latter mainly with regard to the crossover between special and
ordinary transition. For the sake of simplicity,
all calculations are restricted to $\tau=0$. In the following we first
give the results for the one-loop contribution to $m(z)$.
They are regularized dimensionally
in $d=4-\epsilon$ dimensions and renormalized by
minimal subtraction. After focussing attention on
the short-distance behavior of the OP, we calculate
the full scaling functions up to first order in
$\epsilon$ near both the ordinary
and the special transition. Finally we present results on the
crossover between special and ordinary near-surface behavior.

In order to remove the UV divergences appearing in the one-loop term,
the parameters $u$, $c$, and $h_1$ of Sec.\,\ref{0loop}
have to be considered as bare parameters labelled with an index `0' below.
In the standard procedure, dimensional regularization combined with
minimal subtraction\cite{diehl}, the bare parameters are replaced by
renormalized ones to obtain a UV finite expression for $m(z)$.
The field $\phi$ which in general also has to be renormalized
is unchanged up to order $\epsilon$, such that we do not have
to worry about wavefunction renormalization in the present work.
Further, in dimensional regularization the fixed point value $c^*$
of the surface enhancement remains at zero such that
$\tilde c =c$.

\subsection{Perturbation Expansion}\label{loop1}
Expanding in terms of the coupling constant $u_0$, $m(z)$
can be written in the form 
   \begin{equation}\label{per}
    m(z)= m^{(0)}(z) +m^{(1)} (z) + O(u_0^2)\>,
    \end{equation}
where the MF solution $m^{(0)}$ was given in (\ref{mMF}) and
the first-order term $m^{(1)}$ is
    \begin{equation}\label{def}
     m^{(1)}(z)= -\frac{u_0}{2}\int_0^{\infty} dz'\, \hat G({\bf 0}, z,z')  
    \int_{\bf p}\hat G({\bf p}, z',z')\, m^{(0)} (z')
    \end{equation}
with $\int_{\bf p}\equiv (2\pi)^{-d+1}\int\,d^{d-1}p$.
Using the results (\ref{mMF}), (\ref{chi}), and (\ref{G(0)}), the
$z'$ integration in (\ref{def}) can be carried out. This yields
the result
\begin{equation}\label{mone}
   m^{(1)} = -\frac{u_0 s_d\sqrt\pi}{z^+\,\Gamma
   \left((3-\epsilon)/2\right)}\alpha^{-1} X\,\int_0^{\infty}
    d k\,   k^{-\epsilon}\, I
   \end{equation}
where $s_d=(2\sqrt{\pi})^{-d}$, and the integrand $I$ is given by
  \begin{eqnarray}\label{In}
\lefteqn{  I= -B\,  F\, e^{-2k\,z/z^+} 
  +\frac{3}{2}\,  X   \left[\left(B+1\right)\left(1+
  \frac{1}{q}\right) -X\right] k^{-3}+\frac{1}{2q} X B} 
 \nonumber \\[3mm]
 & &\hspace*{12mm} 
+\frac{1}{4q}\left[ B\,X (q+9)+3X(q-1)-3q\right] k^{-1}\nonumber\\[3mm]
 & &\hspace*{12mm}
+ \left( \frac{X^2}{2q} 
+\frac{1}{6X}  -\frac{X^2}{6} +B\, X \left( \frac{5}{4} + \frac{3}{q}\right) k^{-2}\right) k\>,
  \end{eqnarray}
where $q$ and $k$ were defined in (\ref{q}) and (\ref{k}),
respectively.
Further, we introduced
\begin{equation}\label{X}
X=\frac{z^+}{z+z^+}
\end{equation}
with $z^+$ defined in (\ref{z0}) and
\begin{equation}\label{lop5}
F = \frac{1}{4} k^{-1} 
+\frac{5}{4} X k^{-2} +\frac {3}{2} X^2 k^{-3}\>.
\end{equation}

\subsection{Renormalization of UV Divergences and Short-Distance Singularities}\label{loop2}
The perturbative result (\ref{mone}) contains UV singularities
which show up as poles $\propto \epsilon^{-1}$ in dimensional
regularization. These poles are removed
by minimal subtraction\cite{diehl,smock}, by expressing the
bare parameters occurring in $m^{(0)}$ in terms of renormalized ones.
The relation between bare and renormalized coupling is given by
\begin{equation}\label{renu}
u_0\,s_d=\mu^{\epsilon}\,u\left[ 1+\frac{3 u}{\epsilon}+{\cal O}(u^2)
\right]\>,
\end{equation}
where $\mu$ is an arbitrary momentum set to $1$ afterwards.
Further, near the special transition, for
finite $c$ and $h_1$, the renormalizations are
\begin{equation}\label{rensp1}
c_0=\mu c \left[ 1+\frac{u}{\epsilon}+{\cal O}(u^2)\right]
\end{equation}
and
\begin{equation}\label{rensp2}
h_{1,0}=\mu^{d/2} h_{1} \left[ 1+\frac{u}{2 \epsilon}+{\cal O}(u^2)\right]\>.
\end{equation}
At the ordinary transition, for $c\to \infty$ with finite
${\sf h}_1=h_1/c$, we have
\begin{equation}\label{renord}
{\sf h}_{1,0}=\mu^{d/2} {\sf h}_{1} \left[ 1+\frac{u}{2 \epsilon}+{\cal O}(u^2)\right]\>.
\end{equation}
Up to first order in $\epsilon$ the renormalization of $h_1$ and
${\sf h}_1$ are the same, but taking into account higher orders
in $\epsilon$ they are different\cite{diehl}. In both cases the poles
are removed and one is left with an expression for $m(z)$ that
is regular in $\epsilon$ and $u$. 

In our first-order perturbative calculation, modifications of the
short-distance behavior of the OP profile show up in the form of logarithmic
(short-distance) singularities of $m^{(1)}$. The source
of these logarithms in our result (\ref{mone}) is the first term
on the right-hand side of (\ref{In}).
Focussing the attention for the moment on the short-distance
behavior only, (\ref{mone}) can be written as
   \begin{equation}\label{sds}
   m(z)=\frac{u\,s_d \,X}{2\, z^+ \alpha} J_{sing} + \mbox{terms regular for
$z\to 0$}\>,
   \end{equation}
where $J_{sing}$ is given by
\begin{equation}\label{jsing}
J_{sing}=\int_1^{\infty} dk\, B\,k^{-1-\epsilon}\,e^{-2k\,z/z^+}\>.
\end{equation}
For a closer inspection of the $z\to 0$ limit of this integral,
it is useful to rewrite the function $B$ (defined in (\ref{BB})) in
the equivalent forms
\begin{equation}\label{B}
  B=1-2b_{sp}=-1+ 2k b_{ord}
\end{equation}
with
\begin{equation}\label{bs}
      b_{sp}=\frac{(q+1)k^2+3q}{k^3 +(q+1)k^2+3qk +3q}
\end{equation}
and
\begin{equation}\label{bo}
  b_{ord}=\frac{k^2+3q}{k^3 +(q+1)k^2+3qk +3q}\>.
\end{equation}
While $b_{sp}$ yields regular terms for general $c$ and $h_1$, in the second
form $b_{ord}$ leads to finite contributions after taking the limit $c\to \infty$:
 \begin{equation}\label{binf}
     \lim_{c\to \infty} b_{ord}\equiv b_{\infty} =\frac{3}{k^2 +3k +3}\>.
    \end{equation}
With (\ref{bs}) and (\ref{bo}), $J_{sing}$ can be written as
\begin{equation}\label{cross}
J_{sing}\! =\!\! \int_1^{cz^+}\!\! d k 
   (2kb_{ord}-1)k^{-1-\epsilon}  e^{-2kz/z^+}
  \! \!+\!\!\int_{cz^+}^{\infty}\!\! d k (1 -2b_{sp})
   k^{-1-\epsilon}   e^{-2kz/z^+}\!\!,
\end{equation}
where we assumed that $c z^+>1$. Alternatively,
after straightforward manipulations, it can be expressed in the form 
\begin{equation}\label{cross2}
   J_{sing}\! =\! \int_1^{\infty} d k k^{-1-\epsilon} \!
   \left( 2e^{-2kz/l_c}   -e^{-2k z/z^+}\right)\!+\!
   \mbox{terms regular for $z \to 0$ }\!,
\end{equation}
where $l_c\equiv c^{-1}$ is the length scale induced by the surface
enhancement.

To discuss the crossover between
special and ordinary near-surface behavior, we assume $c$ to be large
and  $l_c \ll z^+\simeq l_{ord}$. From (\ref{cross2}) it becomes
then apparent that for $z\ll l_c$
\begin{equation}\label{sdss}
     J_{sing}=\log z 
     + \mbox{terms regular for $z\to 0$ } \>.
     \end{equation}
For $l_c\ll z\ll z^+$, on the other hand, the first term in the integral in (\ref{cross2}) is negligible, and we obtain
   \begin{equation}\label{sdso}
   J_{sing}=-\log z +\mbox{terms regular for $z\to 0$ }\>,
\end{equation}
which also holds for $c\to \infty$ and $z\ll z^+$, i.e., it
describes the short-distance singularity at the ordinary transition.

If eventually the coupling constant $u$ is set to its fixed-point value
$u^*=\epsilon/3$ and the logarithms are exponentiated,
we find
 \begin{equation}\label{sdp}
      m(z)\sim\left\{ \begin{array}{ll}
      1-\frac{\epsilon}{6}\log z +O(\epsilon ^2)
      \sim z^{-\epsilon/6} &   \mbox{for $z\ll l_c$}\\[4mm]
      1+\frac{\epsilon}{6}\log z +O(\epsilon ^2)
      \sim z^{\epsilon/6} &   \mbox{for $l_c\ll z\ll l_{ord}$}
      \end{array}
      \right. \>.
      \end{equation}
The exponents $\epsilon/6$ and $-\epsilon/6$ are to be
identified with the first-order values for $1-\eta_{\perp}^{ord}$ and
$(\beta_1^{sp}-\beta)/\nu$, respectively \cite{diehl}.
In general, for finite $c z^+$ both terms in (\ref{cross})
give non-negligible contributions, each on the respective
length scale, i.e. the first term only for $z\ll l_c$,
the second one for $z\ll l_{ord}$. This is consistent with
our qualitative discussion in Sec.\,\ref{crospord}.

\subsection{Scaling Function at the Ordinary Transition.}\label{loop3}
 
The procedures for obtaining the scaling function
at the ordinary and the special transition are similar.
We describe the first case in more detail. 
For $c\to \infty$, after the subtraction of
pole terms with (\ref{renu}) and (\ref{renord}), the perturbative
results (\ref{per}) and (\ref{mone}) take the form
    \begin{equation}\label{ror}
    m(z)=m^{0}(z;u,{\sf h}_1)-\frac{2us_dX}{z^+\alpha}
    \left[\int_0^{\infty} dk\, C^{ord} +\frac{1}{4} 
    \left( 2X-3\right)\log  z^+\right]\>,
    \end{equation}
where    
  \begin{eqnarray}\label{Cor}
  C^{ord}  & = & {F}(1-b_{\infty}) e^{-2kz/{z^+}} 
  +\frac{3}{2}   X(b_{\infty} -X) k^{-3} +\frac{5}{4}  X 
  (b_{\infty} -1) k^{-2} \nonumber \\
  & + & \frac{1}{4} (2X-3) \frac{1}{k(k+1)} + \frac{1}{4} b_{\infty} X k^{-1}
  \end{eqnarray}
is obtained from $I$ (defined in (\ref{In}))
by setting $q=c=\infty$. 
In dimensional regularization terms of order $O(k^0)$ and $O(k^1)$ present in (\ref{In}) give vanishing contributions to the integral and are omitted here.
As the next step we utilize the scaling form
(\ref{scalordm}), derived on a more rigorous level with the help
of the renormalization group \cite{diehl}.
The scaling function is given by
        \begin{equation}\label{rgo1}
        {\cal M}_{ord} (\zeta) = 
        m(z=1,\bar u=u^*,
\bar{\sf h}_1(\zeta)={\sf h}_1   
        z^{\Delta_1^{ord}/\nu})\>,
        \end{equation}
where the scaled distance $\zeta$ is
  \begin{equation}\label{Z}
  {\zeta}= \frac{z}{l_{ord}}\qquad \mbox{with}\qquad l_{ord}
=\left(\alpha^*\,{\sf h}_1\right)^{-\nu/\Delta_1^{ord}}\>,
  \end{equation}
and $\alpha^*$ is defined by (\ref{alpha}) with $u=u^*$. 

The result (\ref{ror}) gives the
scaling function up to terms $O(\epsilon^2)$ according to the equation
   \begin{equation}\label{scfo}
   {\cal M}_{ord} (\zeta) =   
   m^{(0)}(1,u^*,{\sf h}_1    
   z^{\Delta_1^{ord}/\nu} )+   
   m^{(1)}(1,u^*,{\sf h}_1
    z^{\Delta_1^{ord}/\nu})  +O(\epsilon^2)\>
   \end{equation}
The MF result (\ref{mMF}), after substituting  $z\to 1$ and 
   \begin{equation}\label{z+}
    z^+\to (\alpha^* {\sf h}_1)^{-1} z^{-\Delta_1^{ord}/\nu} =\zeta ^{-1} 
   \left( 1 +   \frac{\epsilon}{3} \log \zeta \right) +O(\epsilon ^2)
   \end{equation}
reads
     \begin{eqnarray}\label{MFo}
      m^{(0)}(z=1,\bar u=u^*,\bar {\sf h}_1 =
{\sf h}_1 z^{\Delta_1^{ord}/\nu} ) &=&\nonumber\\[3mm]
  & &\hspace*{-3cm}    (\alpha^*)^{-1}\frac{\zeta}{1+\zeta} \left( 1-
\frac{\epsilon}{3} \frac{1}{1+\zeta} 
      \log \,\zeta \right) +O(\epsilon ^2)\>.
      \end{eqnarray}
Together with the one-loop contribution (\ref{ror}),
we obtain for the scaling function
   \begin{equation}\label{sfo}
   {\cal M}_{ord} (\zeta)= (\alpha^*)^{-1}\frac{\zeta}{1+\zeta} \left[ 1-
   \epsilon\,\left(  \frac{1}{2} \log\zeta + \frac{2}{3} 
   \int_{0}^{\infty} d k\, C^{ord}\right) \right] +
   {\cal O}(\epsilon ^2)\>.
   \end{equation}
One can easily check that for $\zeta\to 0$ the integral
$\int_0^{\infty} dk\, C^{ord}$
behaves as $ \sim -\frac{1}{4} \log \zeta + $
regular terms, and, as a consequence, the short-distance
limit of the scaling function is given by
    \begin{equation}\label{sfos}
    {\cal M}_{ord} (\zeta)\sim  \zeta \left[ 1- 
    \frac{\epsilon}{3} \log \zeta \right] +
{\cal O}(\epsilon^2)\>.
    \end{equation}
Taking into account that $\beta/\nu=1-\epsilon/2+{\cal O}(\epsilon^2)$
such that $z^{-\beta/\nu}=z^{-1}(1+\epsilon/2\,\log z
+{\cal O}(\epsilon^2)$), this is
consistent with (\ref{sdp})
and, hence, with the result (\ref{power}) of our scaling analysis.
Furthermore, it is straightforward to show that ${\cal M}_{ord}$
approaches a constant for $\zeta\to \infty$. 

\begin{figure}[b]
\def\epsfsize#1#2{0.6#1}
\hspace*{1.7cm}\epsfbox{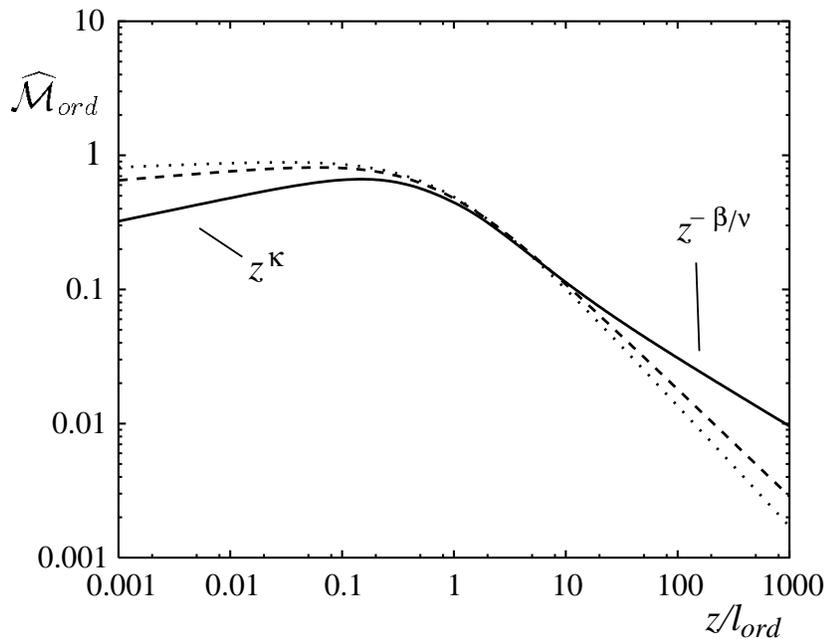}
\caption{Scaling function $\widehat{\cal M}_{ord}$ for
the crossover between ordinary and normal transtion defined
in (4.32) for $\epsilon=1$ (solid), 0.4 (dashed), and
0.1 (dotted line).}
\end{figure}
  
In order to exponentiate the scaling function in the
short-distance regime, we effectively replace the
logarithm of (\ref{sfos}) by the power law $\zeta^{1-\epsilon/3}$,
the exponent $1-\epsilon/3$ being the one-loop value of $\Delta_1^{ord} /\nu$.
This is achieved by adding 
   \begin{equation}\label{Ginf}
  {\cal G}_{ord}= e^{-\zeta^2} \left[\zeta^{1-\epsilon/3}  
   -\zeta \left(1-\frac{\epsilon}{3} \log \zeta\right)\right],
   \end{equation}
to (\ref{sfo}), a function of ${\cal O}(\epsilon^2)$.

The result for the OP as a function of $\zeta$, i.e. 
the function 
\begin{equation}\label{hatMord}
\widehat {\cal M}_{ord}=\zeta^{-1+\epsilon/2}
({\cal G}_{ord}+\alpha^*{\cal M}_{ord})\>,
\end{equation}
is depicted in double-logarithmic representation for three different
values of $\epsilon$ in Fig.\,4. The integral in (\ref{sfo}) had
to be calculated numerically.
All profiles show the characteristic crossover
between short-distance growth and long-distance decay. However, extrapolating
$\epsilon$ to one obviously leaves us with a profile with somewhat
unexpected features.
It has a point of inflection were one actually expects a smoother
scaling function. On the other hand, for the
smaller values of $\epsilon$ this problem does not exist. So we
think that extrapolating $\epsilon$ to 1 in
(\ref{sfo}) probably just at the limit,
where the first-order result ceases to provide a good approximation.

\subsection{Scaling Function at the Special Transition and
Crossover to the Ordinary Transition}\label{loop4}

For arbitrary $c$ we proceed as before, now with 
the renormalization of the surface parameters given by (\ref{rensp1})
and (\ref{rensp2}). 
The perturbative result for the profile (\ref{per}) is \cite{ciach}
  \begin{equation}\label{rsp}
  m(z)=m^{(0)}(z;u,c,h_1)\! - \!\frac{2us_dX}{z^+ \alpha}\left[ 
  \int_0^{\infty} dk\, C(k)\! +\! \frac{1}{2} 
  \left(\frac{X}{q}  -\frac{3}{2}\right)\log z^+\right]\!,
  \end{equation}
where we extracted the regular contribution   
\begin{eqnarray}\label{C}
    C & = & (2b_{sp}-1)\,F\, e^{-2k z/{z^+}} 
    + \frac{3}{2} X \left[2(1-b_{sp})\left(1+
    \frac{1}{q}\right)  -X\right]k^{-3}\nonumber  \\
   &- &\frac{(q+9)X}{2q}b_{sp} k^{-1} 
     + \left(\frac{5}{4} +\frac{3}{q}\right)(1-2b_{sp})
    X\,k^{-2} \nonumber  \\
& +& \frac{1}{2}\left(\frac{X}{q} -
    \frac{3}{2}\right)\frac{1}{k(k+1)}
   - \left(b_{sp}-\frac{q+1}{k}\right)\frac{X}{q}
    \end{eqnarray}
from $I$ of (\ref{In}).

At the special transition, alternatively to (\ref{scalspm}), the order parameter can be expressed in the scaling form 
  \begin{equation}\label{rgsp}
  m(z;u,\bar c,\bar h_1) = z^{-\beta/\nu}
  {\cal M}(\zeta,\gamma)
  \end{equation}
with
  \begin{eqnarray}\label{sfst}
   {\cal M}(\zeta,\gamma)  & = & m( z=1; \bar u=u^*, 
  \bar c(\zeta,\gamma)=c z^{\phi/\nu},\bar h_1(\zeta) = h_1 z^{\Delta_1^{sp}/\nu})\nonumber \\[3mm]
  & & \hspace*{-1.6cm}= m^{(0)}(1;u^*,cz^{\Phi/\nu},h_1z^{\Delta_1^{sp}/\nu}) 
  +  m^{(1)}(1;u^*,cz^{\Phi/\nu},h_1z^{\Delta_1^{sp}/\nu}) + O(\epsilon ^2)
  \end{eqnarray}
where
   \begin{equation}\label{zetagamma}
   \zeta=\frac{z}{l_{sp}}\>,\quad \gamma = ch_1^{-\Phi/\Delta_1^{sp}}
\>, \quad \mbox{and}\quad l_{sp}=\left(\alpha^*\, h_1\right)^{-\nu/\Delta_1^{sp}}\>.
   \end{equation}
Expressing all the parameters in terms of scaling variables $\zeta$ and $\gamma$ and keeping terms to first order in $\epsilon$ in the
MF contribution and to zeroth order in $\epsilon$ in the one-loop term,
we obtain the result
    \begin{eqnarray}\label{M}
{\cal M}(\zeta,\gamma)&=&(\alpha^*)^{-1} \frac{\zeta}{\zeta+\zeta_0} 
    \left\{ 1  - \frac{\epsilon}{3}\left[\frac{1}{q} \frac{\zeta_0}{\zeta+\zeta_0}\log\zeta_0 \right.\right.
\nonumber\\[3mm]
& & \hspace*{2cm} +  \left.\left. \frac{3}{2} \log\left(\frac{\zeta}{\zeta_0}\right)
    +2 \int_0^{\infty}dk\, C(k)\right]\right\} +O(\epsilon ^2)\>,
    \end{eqnarray}
where
  \begin{equation}\label{zeta0}
  \zeta_0=\frac{1}{2}\left(\gamma + \sqrt{\gamma^2+4}\right)
  \end{equation}
and the integrand $C$ is given by (\ref{C}) with $z= 1$, $z^+= \zeta_0/\zeta$, and $q=
2+\zeta_0\gamma$. 

It is straightforward to verify that the short-distance behavior
of the OP described by (\ref{M}) for a system
with $\gamma \gg 1$ is consistent with the
discussion in Sec.\,\ref{loop2}, especially with (\ref{sdp}).
The short-distance logarithm in (\ref{M}) is caused by the
first term on the right-hand side of (\ref{C}). One obtains
\begin{equation}\label{Mm}
   {\cal M} (\zeta,\gamma) \sim \frac{\zeta}{\zeta_0} 
   \left[ 1-\frac{2}{3}    \epsilon \log
   \left(   \frac{\zeta}{\zeta_0}\right)\right] + 
   {\cal O}(\epsilon ^2)
\end{equation}
for $\zeta/\zeta_0\ll q^{-1}$ (equivalent to $z\ll l_c$), which,
together with the prefactor $\sim \zeta^{-\beta/\nu}$ (see (\ref{rgsp})),
is consistent with the upper line in (\ref{sdp}) and with
Refs.\,\cite{brezin,ciach}.
And one finds
\begin{equation}\label{MM}
  {\cal M} (\zeta,\gamma) \sim \frac{\zeta}{\zeta_0} 
  \left[ 1-\frac{1}{3}   \epsilon 
  \log\left(  \frac{\zeta}{\zeta_0}\right)\right] + {\cal O}(\epsilon ^2)
\end{equation}
for $q^{-1}\ll \zeta/\zeta_0\ll 1$ (equivalent to $l_c\ll z\ll l_{ord}$), 
in agreement with the second case in (\ref{sdp}). 

The crossover between the two asymptotic cases occurs at the length scale $l_{sp}$, which is the intermediate scale between $l_c$ and $l_{ord}$. The ratio between all the scales is set by the single parameter $\gamma$
of (\ref{zetagamma}),
and from the definitions of the length scales (\ref{lengthsp}), (\ref{lengthord}) 
and $\gamma$ we find
\begin{equation}\label{lengths}
\frac{l^{ord}}{l^{sp}}=\gamma^{y\nu/{\Delta_1^{ord}}}
\qquad\mbox{and}\qquad
\frac{l^{sp}}{l_c}=\gamma^{\nu/\phi}\>.
\end{equation}

At the special transition, $\gamma =0$, the scaling function is given by
${\cal M}_{sp}={\cal M}(\zeta,\gamma=0)$, with ${\cal M}$
given by  (\ref{M}). Like at the ordinary transition we
exponentiate the short-distance logarithm by adding
the ${\cal O}(\epsilon^2)$ function 
  \begin{equation}\label{G}
  {\cal  G}_{sp}= e^{-\zeta^2} \left[\zeta^{1-2\epsilon/3}  -\zeta 
   \left(1-\frac{2}{3}   \epsilon\log \zeta\right)\right]\>,
   \end{equation}
where the exponent $1-2\epsilon/3$ is the first-order value for $\beta_1^{sp}/\nu$. 
The OP scaling function
\begin{equation}\label{hatMsp}
\widehat {\cal M}_{sp}= \zeta^{-1+\epsilon/2}\left({\cal G}_{sp}+
\alpha^*{\cal M}_{sp}\right)\
\end{equation}
is depicted in Fig.\,5 for three values of $\epsilon$. All
profiles correctly show the asymptotic power laws for short
and long distances, respectively. Like the result for the ordinary
transition, however, the curve for $\epsilon=1$ exhibits
unexpected features. This becomes clearer visible, when one
plots the scaling function ${\cal M}_{sp}$ directly, without
exponentiation, i.e. the result (\ref{M}) with $\gamma=1$.
The result is shown in Fig.\,6 (upper curve).
For $z\to 0$, ${\cal M}_{sp}$ behaves as $\sim \log \zeta$, and
for $\zeta\to \infty$ it approaches a constant. As seen from Fig.\,6,
it varies between the two asymptotic limits in a non-monotonic
fashion, where one actually expects a monotonic variation.
The reason for this qualitative defect was already discussed at
the end of Sec.\,\ref{loop3}.

\begin{figure}[b]
\def\epsfsize#1#2{0.6#1}
\hspace*{2cm}\epsfbox{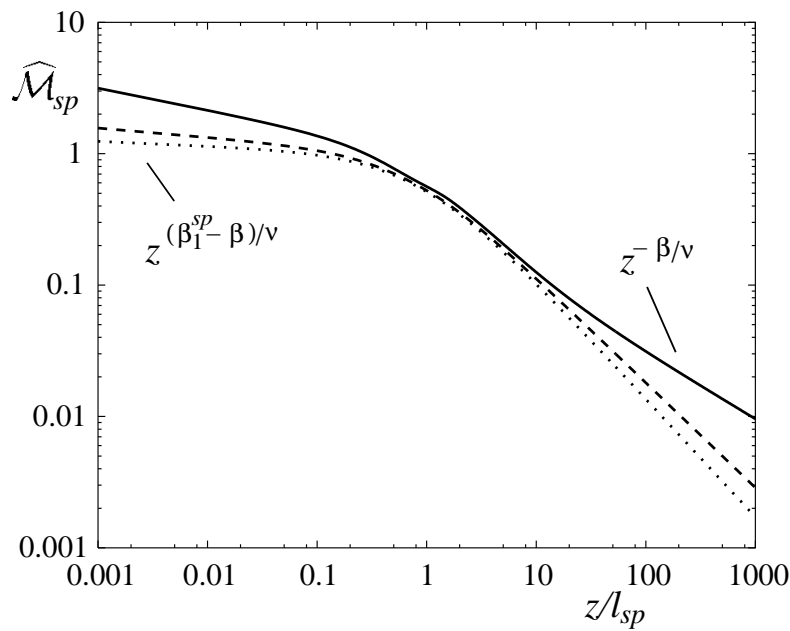}
\caption{Scaling function $\widehat{\cal M}_{sp}$ for
the crossover between special and normal transition defined
in (4.44) for $\epsilon=1$ (solid), 0.4 (dashed), and
0.1 (dotted line).}
\end{figure}

What happens if $\gamma$ takes intermediate values, between
the fixed points $\gamma=0$ (special) and $\gamma=\infty$ (ordinary)?
The results for ${\cal M}(\zeta,\gamma)$ of (\ref{M})
for $\gamma=0$, 1, 5, 10, and 15 and 
$\epsilon=1$ are displayed in Fig.\,6
(from top to bottom). 
For increasing $\gamma$ the profiles first become monotonic
functions and then, for $\gamma\gtrsim 1$, a minimum occurs. 
This is the signature of the
crossover between special and ordinary transition. Consistent
with the qualitative discussion in Sec.\,\ref{crospord},
drops first and then increases again to approach
a constant. However,
for larger values $\gamma\gtrsim 15$ the scaling function ${\cal M}$
becomes negative, signaling that the one-loop result
calculated at the special transition becomes meaningless
at intermediate values of $\zeta$. This time the problem is not
cured by going to smaller $\epsilon$;
for large $\gamma$ the
scaling function ${\cal M}$ becomes still negative.

\begin{figure}[t]
\def\epsfsize#1#2{0.6#1}
\hspace*{2cm}\epsfbox{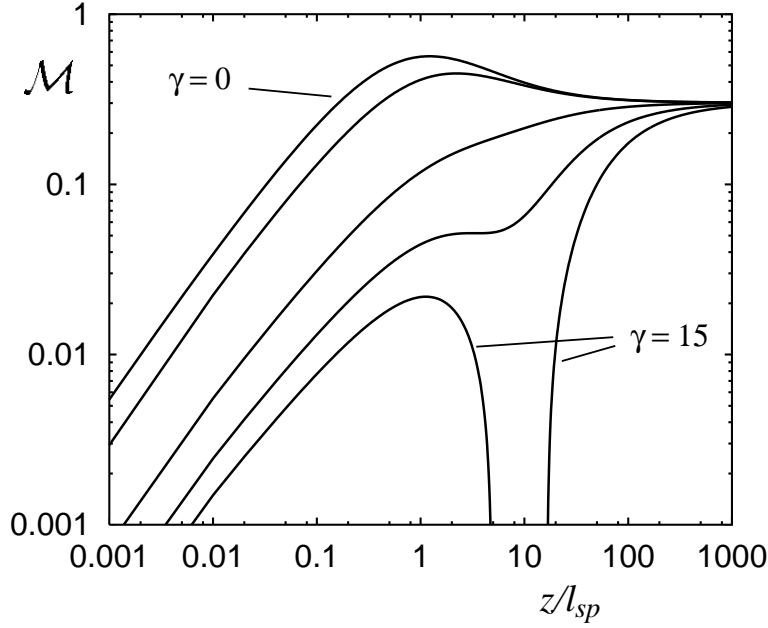}
\caption{The (unexponeniated) scaling function ${\cal M}$
of (4.38) for different values of $\gamma$ and $\epsilon=1$.}
\end{figure}

\begin{figure}[t]
\def\epsfsize#1#2{0.6#1}
\hspace*{2cm}\epsfbox{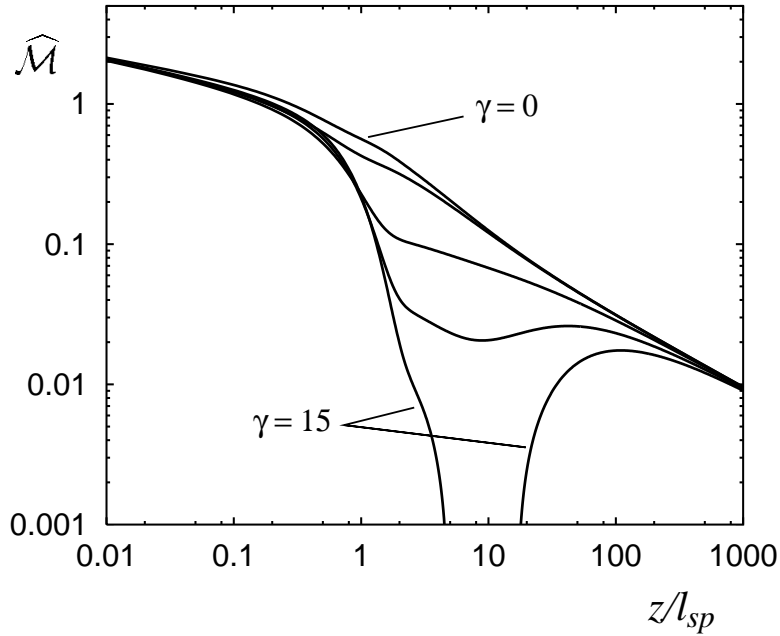}
\caption{Scaling function $\widehat{\cal M}$ for
the order parameter in the crossover regime
between special and normal transtion defined
in (4.45) for different values of $\gamma$ and $\epsilon=1$.}
\end{figure}

Eventually in Fig.\,7 we have plotted the function
\begin{equation}\label{hatM}
\widehat {\cal M}= \zeta^{-1+\epsilon/2}\left({\cal G}_{sp}+
\alpha^*{\cal M}\right)\>,
\end{equation}
where ${\cal G}_{sp}$ is defined in (\ref{G}), for the same
values of $\gamma$ as in Fig.\,6. All curves have the expected
power-law form for small and large arguments. For intermediate
values up to $\gamma \simeq 10$, the behavior is reminiscent to
the qualitative picture described in Sec.\,\ref{crospord} and
depicted in Fig.\,1. For larger $\gamma\gtrsim 15$
the OP becomes negative for intermediate $z/l_{sp}$.

\section{Summary and Concluding Remarks}\label{summary}
We studied a semi-infinite Ising-like system by means of
continuum field theory. Particular attention was paid to
situations where the surface enhancement $c$ and the magnetic
field $h_1$ are off their fixed-point values, in which case
they may give rise to macroscopic length scales (in addition
to $\xi$) and anomalous behavior of thermodynamic quantities
near the surface. 
We carried out a phenomenological scaling analysis
and discussed the near-surface phenomena from various points of
view, especially analogies with in critical dynamics were pointed
out. Furthermore, we presented mean-field results for the
structure function in the crossover regime between ordinary
and normal transition, and we carried out a first-order calculation
for the order-parameter in the framework of renormalization-group
perturbation theory ($\epsilon$-expansion).

Let us conclude by highlighting the main
results of this work. The structure functions displayed in Fig.\,3
may, on a qualitative level, directly be compared to the
outcome of scattering experiments, like the one carried
out by Mail\"ander et al. \cite{mail}. Our results show that
in the case when the length $l_{ord}$ is larger than the penetration depth of the radiation $\kappa_i^{-1}$, the structure function is effectively 
governed by the exponent of the ordinary transition
and shows the cusp for $p\to 0$ predicted in Ref.\,\cite{diwa}.
Only when the product $l_{ord}\cdot\kappa_i$ is smaller, the signature
of the normal transition, a ``flat'' structure function, can be seen.

Further, the order-parameter profiles for the crossover
between ordinary and normal transition displayed in Fig.\,4 supplement
earlier studies which focussed on various other
aspects of the near-surface behavior\cite{czeri,swewa,alina}.
Although extrapolating $\epsilon$ to one
in our results appeared to be rather courageous, we believe
that with these profiles a quantitative comparison with experimental
data should be feasible. 
The Monte Carlo data available at present\cite{swewa},
although obtained from relatively large
systems with up to $256^2\times 512$ spins,
are still distorted by large finite-size
effects and can not yet compete with the results presented
in this work, where really the situation in the semi-infinite system is
described.

Eventually, the crossover between special and ordinary transition
was studied on the basis of the first-order perturbative approximation.
In principle the results for general
$c$ and $h_1$ presented in Sec.\,\ref{loop4} should cover
this crossover as the limit $c\to\infty$. The qualitative scenario
was described in Sec.\,\ref{crospord}.
In our first-order calculation we found that, while the
asymptotics for small and large $z$ remain correct
(compare also (\ref{sdp}),
the one-loop contribution becomes larger than the zero-loop term
for intermediate $z$, and the order-parameter
profile becomes even negative for large $c$ (see Figs.\,6 and 7). This means,
in other words, that the first-order approximation breaks down
in this limit, at least for intermediate distances,
where the crossover from $z^{(\beta_1^{sp}-\beta)/\nu}$ (characteristic
for the special transition) to $z^{(\Delta_1^{ord}-\beta)/\nu}$ (characteristic
for the ordinary transition) should occur. The reason for the failure
of the first-order approximation is connected to different
UV divergences at the ordinary and special transition, which,
in the framework of the $\epsilon$-expansion, are intimately
related to the short-distance singularities. This
problem could probably be solved by introducing a cutoff, which, from the
physical point of view, would certainly be meaningful, and consider
the ordinary transition as the case where the surface enhancement $c$
is of the order of the cutoff (as opposed to being of the order
of the finite momentum scale $\mu$ in dimensional regularization).\\[5mm]
{\small {\it Acknowledgements}:  We thank H.\ W.\ Diehl for discussions and comments. A.\,C. is particularly grateful for very interesting discussions and helpful suggestions during the early stages of this work.
This work was supported in part by the Deutsche Forschungsgemeinschaft
through Sonderforschungsbereich 237 and in part by the KBN grants No. 2P30302007 and No. 2P03301810.}
\newpage

\end{document}